\documentclass[useAMS,usenatbib]{mnras}

\usepackage{tabularx,ragged2e,booktabs,caption}
\usepackage{amsmath}
\usepackage{amssymb}
\usepackage{epsfig}
\usepackage{ifthen}
\usepackage{latexsym}
\usepackage{rotating}
\usepackage{times,epsf}
\usepackage{txfonts}
\usepackage{varioref}
\usepackage{verbatim}
\usepackage{array}
\usepackage{url}
\usepackage{color}
\usepackage[dvipsnames]{xcolor}
\usepackage[T1]{fontenc}
\usepackage{epstopdf}
\usepackage{ulem}
\usepackage{graphicx}

\newcommand{\be}{\begin{equation}}
\newcommand{\ee}{\end{equation}}
\newcommand{\bea}{\begin{eqnarray}}
\newcommand{\eea}{\end{eqnarray}}

\newcommand{\Medd}{\dot M_{\rm Edd}}
\newcommand{\Rg}{R_{\rm g}}
\newcommand{\rg}{R_{\rm g}}
\newcommand{\tg}{t_{\rm g}}

\title[Kinetic and radiative efficiencies in accretion flows]{Kinetic and radiative power from optically thin accretion flows}
\author[A. S\k{a}dowski \& M. Gaspari]
       {Aleksander S\k{a}dowski$^{1,}$\thanks{E-mail: asadowsk@mit.edu (AS); Einstein Fellow} \&  Massimo Gaspari$^{2,}$\thanks{E-mail: mgaspari@astro.princeton.edu (MG); Einstein \& Spitzer Fellow}\\
 $^1$ MIT Kavli Institute for Astrophysics and Space Research, 77 Massachusetts Ave, Cambridge, MA 02139, USA\\
$^2$ Department of Astrophysical Sciences, Princeton University, 4 Ivy Lane, Princeton, NJ 08544, USA
}

\begin{document}

\maketitle

\label{firstpage}

\begin{abstract}
  We perform a set of general relativistic, radiative,
  magneto-hydrodynamical simulations (GR-RMHD) to study the transition
  from radiatively inefficient to efficient state of accretion on a
  non-rotating black hole.  We study ion to electron temperature
  ratios ranging from $T_{\rm i}/T_{\rm e}=10$ to $100$, and simulate
  flows corresponding to accretion rates as low as $10^{-6}\Medd$, and
  as high as $10^{-2}\Medd$. We have found that the radiative output
  of accretion flows increases with accretion rate, and that the
  transition occurs earlier for hotter electrons (lower $T_{\rm
    i}/T_{\rm e}$ ratio). At the same time, the mechanical efficiency
  hardly changes and accounts to ${\approx}\,3\%$ of the accreted rest
  mass energy flux, even at the highest simulated accretion rates. This is
  particularly important for the mechanical AGN feedback regulating
  massive galaxies, groups, and clusters. Comparison with recent
  observations of radiative and mechanical AGN luminosities suggests
  that the ion to electron temperature ratio in the inner,
  collisionless accretion flow should fall within $10<T_{\rm i}/T_{\rm
    e}<30$, i.e., the electron temperature should be several percent
  of the ion temperature.
%is no less than $3\%$, and no more than $10\%$ of the ion temperature.

\end{abstract}

\begin{keywords}
  accretion, accretion discs -- black hole physics -- relativistic
  processes -- methods: numerical
\end{keywords}

\section{Introduction}\label{s.intro}
\label{s.introduction}

Accreting black holes (BH) are behind some of most energetic and fundamental phenomena in the Universe. In the supermassive regime (SMBH), they are often known as active galactic nuclei (AGN) and quasars. When the rate at which gas is lost below the BH horizon is low (normalized to the Eddington unit), accretion flows are optically thin and radiatively inefficient (an extreme case is the accretion flow in the center of the Galaxy). Radiatively inefficient BHs do however inject significant amount of mechanical energy into the environment, preventing catastrophic cooling flows and star formation rates via large-scale outflows, cavities and shocks (\citealp{mcnamara12} for a review), at the same time establishing the celebrated BH mass - velocity dispersion relation (\citealp{kormendy13} for a review). %and often inflating large scale cavities.
The output mode is expected to change when gas is accreted at rates exceeding a percent of the Eddington rate. Under such conditions the previously radiatively inefficient sources become quasars dominated by the radiative output.

The transition from the radiatively inefficient to the efficient
regime is not well understood. Neither the physics of the electron
heating in optically thin collisionless plasmas. On the other hand,
there is ample amount of observational data which probes accreting
systems at very different luminosities and accretion rates.  One key
observational study attempting to constrain the kinetic versus
radiative efficiencies in massive galaxies detected with {\it Chandra}
telescope (mainly brightest cluster galaxies; BCGs) is
\cite{russell+13}. The radiative output is constrained via the nuclear
X-ray luminosity, while the kinetic power is estimated by measuring
the X-ray bubble enthalpy divided by the buoyancy time. The data
corroborates that the mechanical output always dominate at low and
intermediate accretion rates (more in Section \ref{s.constraints}). In
terms of evolution, in the nearby Universe, only a few percent of
systems host a radiatively efficient source (or quasar). In other
words, below redshift $z{\sim}2$, AGN feedback is expected to proceed at
a sub-Eddington rate, i.e., the AGN feedback `maintenance' phase is dominated by kinetic input of energy -- typically via
massive outflows (e.g., see \citealp{gaspari16} for a brief review).
%The same work also remark the rapid variability of AGN.

In this work, we perform state-of-the-art, 3-dimensional (3D) GR-RMHD simulations of accretion flows below and near the transitional regime, and calculate the corresponding mechanical and radiative efficiencies. Moreover, comparing the observed and simulated values of the luminosity at which the radiative efficiency becomes comparable with the mechanical one, we aim to constrain the range of possible values of the electron to ion temperature ratio in accreting systems.

In a parallel work, \cite{gasparisadowski+17}, we apply the mechanical efficiency constrained here, to a model of self-regulated AGN feedback which couples the small (horizon) scale to the large (kpc\,-\,Mpc) scale properties of cool-core systems (as massive galaxies, groups, and clusters). The macro feeding properties are mediated by the recently probed chaotic cold accretion (CCA), i.e., the condensation of multiphase clouds out of the turbulent plasma halo, which rain toward the nuclear region and are efficiently funnelled via chaotic inelastic collisions cancelling angular momentum (see \citealp{gaspari+13,gaspari+15,gaspari+17}).
The proposed unification model provides the BH growth and AGN outflow properties as a function of the macro properties (e.g., hot halo temperature), and can be used as an effective sub-grid scheme in structure formation simulations and analytic studies, without resorting to the fine-tuning of a free efficiency.

The presented work is structured as follows. Below we review the observational work we base our comparison on. In Section~\ref{s.method}, we introduce the numerical method and main assumptions. In Section~\ref{s.results}, we discuss the performed 3D GR-RMHD simulations. In Section~\ref{s.implications}, we discuss the key implications. In Sections~\ref{s.caveats} and \ref{s.summary}, we discuss the caveats and summarize the main results, respectively.

\subsection{Observational constraints}
\label{s.constraints}

Accreting black holes are known to both emit radiation and inject significant amounts of mechanical energy in the surroundings. The latter effect often leads to the inflation of cavities in the medium surrounding the BH, and could happen both on parsec scales (for stellar mass BHs), as well as on kpc scale (for AGN). From the volume and pressure of a given cavity ($E_{\rm cav}=4PV$) and its expansion (sound-crossing) time one may infer the mechanical luminosity of the central source.

Several studies aiming at comparing the radiative and kinetic
properties have been recently performed
\citep[e.g.,][]{kingashley+12,mezcuaprieto-14,hlavacek+15,hogan+15,shin+16}
In this work, we mainly use results by \cite{russell+13}, one of the
most comprehensive investigations based on observations of nuclear
X-ray sources and the corresponding cavities in a large sample of 57
BCGs. These authors have found that the nuclear radiation exceeds the
mechanical energy output of the outflow when the mean accretion rate
rises above a few percent of the Eddington luminosity, corresponding
to the onset of the quasar mode. The two components become comparable
already at $10^{-3}-10^{-2}\,L_{\rm Edd}$ and the radiative component is
very weak, sometimes undetectable, for the lowest accretion and
emission rates \citep[cf.~Fig.~12 in][]{russell+13}.  For all such
reasons, below we use the range $10^{-3}-10^{-2}\, L_{\rm Edd}$ as
reference threshold below which the mechanical luminosity of an
accreting black hole system equals or exceeds the corresponding
radiative output.

\subsection{Units}
\label{s.units}
In this work we adopt the following definition
for the Eddington mass accretion rate,
\be
\label{e.medd}
\Medd = \frac{L_{\rm Edd}}{\eta c^2},
\ee
where $L_{\rm Edd}=4\pi GMm_{\rm p} c/\sigma_{\rm T}=1.25 \times 10^{38}  M/M_{\odot}\,\rm erg/s$ is the 
Eddington luminosity for a~BH of mass $M$, and $\eta$ is the radiative efficiency of a~thin
disk around a~black hole with a~given spin $a_* \equiv a/M$. For a zero-spin BH, $\eta\approx0.057$ \citep{nt73} and
$\Medd = 2.48 \times 10^{18}M/M_{\odot}  \,\rm g/s$. We denote the gravitational radius and time as 
$\rg\equiv GM/c^2$ and $\tg\equiv GM/c^3$, respectively.

\begin{table*}
\begin{center}
\caption{Simulated models}
\label{t.models}
\begin{tabular}{lccccccc}
\hline
\hline
Name & $T_{\rm i}/T_{\rm e}$ & $\dot M/\Medd$ & $L_{\rm rad}/\dot M c^2$ & $L_{\rm tot}/\dot M c^2$ & $L_{\rm kin}/\dot M c^2$ & $H/R$ & $t_{\rm end}/\tg$\\
\hline
\hline
\texttt{f2t10} & $10$ & $9.7\times 10^{-7}$ & $0.0005$ & $0.035$ & $0.034$ & 0.38 & 25000\\
\texttt{f3t10} & $10$ & $1.0\times 10^{-5}$ & $0.0026$ & $0.025$ & $0.022$ & 0.36 &29000\\
\texttt{f4t10} & $10$ & $2.7\times 10^{-4}$ & $0.033$ & $0.065$ & $0.032$ & 0.35 &26000\\
\texttt{f5t10} & $10$ & $2.9\times 10^{-3}$ & $0.033$ & $0.067$ & $0.034$ & 0.20 &27500\\
\texttt{f4t30} & $30$ & $1.9\times 10^{-4}$ & $0.0026$ & $0.033$ & $0.030$ & 0.35 &25000\\
\texttt{f5t30} & $30$ & $4.1\times 10^{-3}$ & $0.017$ & $0.056$ & $0.039$ & 0.20 &28000\\
\texttt{f6t30} & $30$ & $1.6\times 10^{-2}$ & $0.020$ & $0.062$ & $0.042$ & 0.17 &28000\\
\texttt{f5t100} & $100$ & $1.7\times 10^{-3}$ & $0.0016$ & $0.032$ & $0.030$ & 0.38 &20500\\
\texttt{f6t100} & $100$ & $1.0\times 10^{-2}$ & $0.014$ & $0.055$ & $0.041$ & 0.16 &27000\\
\hline
\hline
\multicolumn{6}{l}{$H/R$ - average density scale-height measured at $R=15\rg$. }\\
\multicolumn{6}{l}{Other parameters: $a_*=0.0$, resolution:
  336x336x32, $\pi/2$ wedge in azimuth}\\
\end{tabular}
\end{center}
\end{table*}

\section{Numerical method}
\label{s.method}

For the purpose of simulating optically thin and marginally optically
thin accretion flows we adopt the general relativistic, radiative
magneto-hydrodynamical solver \texttt{KORAL} \citep{sadowski+koral,
  sadowski+koral2} which is capable of evolving gas and radiation in
parallel for arbitrary optical depths. The most recent version of
\texttt{KORAL} is capable of evolving two-temperature plasma
\citep{sadowski+electrons}. This approach, however, is not free from
arbitrary choices and numerical issues (see Discussion section
below). For this reason we decided to solve a simplified problem
where we evolve only the mean temperature of the gas (mixture of electrons and ions), 
but for the purpose of calculating radiative
emission, we split the gas temperature into electron and ion
temperatures, $T_{\rm e}$ and $T_{\rm i}$, respectively, following the
prescribed ratio $T_{\rm e}$ and $T_{\rm i}$. For pure hydrogen gas
the species temperature satisfy, \be T_{\rm gas}=1/2\left(T_{\rm
    e}+T_{\rm i}\right).  \ee We account for free-free and synchrotron
emission and absorption, as well as Compton heating and cooling. The exact formulae
for the opacities are given in \cite{sadowski+electrons}. In this work we
use only grey opacities and we do not calculate electromagnetic spectrum 
of generated radiation.

\subsection{Problem  setup}
\label{s.setup}

We performed a set of 9 three-dimensional simulations of accretion
flows on a non-rotating BH. The simulations cover accretion rates from
$10^{-6}$ to $10^{-2}\Medd$ - enough to cover both the low-luminosity
limit and the observed transition to radiatively-efficient
mode\footnote{Although tempting, simulating accretion flows with
  accretion rates higher than $10^{-2}\Medd$ is extremely challenging
  because such flows collapse to a geometrically thin disk which
  accretes very slowly, and, therefore, obtaining converged solution
  in a reasonable volume requires both extreme numerical resolution
  and long simulation time.}.  We tested three values of ion to
electron temperature ratios, $T_{\rm i}/T_{\rm e}=10$, $30$ and
$100$. This choice reflects the fact that the expected electron
temperature is lower than the ion's (only electrons cool emitting
photons), and, as discussed below, is wide enough to put limits on the
temperature ratio consistent with observations.

Initially, we
performed a single, non-radiative, simulation with an equilibrium
torus \citep[same as in][]{narayan+adafs} threaded by a magnetic field
forming set of poloidal loops \citep[again, same as in][but flipping
the polarity also across the equatorial plane]{narayan+adafs} on a
grid of 336 cells in radius and polar angle, and 32 cells in azimuth
spanning $\pi/2$ wedge. Such a setup was evolved for $15000\,\tg$,
long enough for establishing a now-standard, scale-free solution of a thick
and hot accretion flow with converged region extending up to $\sim
25\rg$.

We then used the final state of this non-radiative simulation as a
starting point for the radiative runs being the subject of this
paper. When starting a radiative run we rescaled the density to a
desired value (which ultimately determined the final accretion rate)
keeping the magnetic to gas pressure ratio and the gas temperature
fixed. At the same time we turned on the radiative evolution allowing
for emission and absorption of radiation with efficiency determined by
the choice of species temperature ratio. In such a way we simulated
the 9 models described in this work until the final time $t_{\rm end}$
(Table~\ref{t.models}), typically $25000\,\tg$.

\section{Results}
\label{s.results}

\subsection{General properties}

 \begin{figure*}
  \includegraphics[width=1.5\columnwidth]{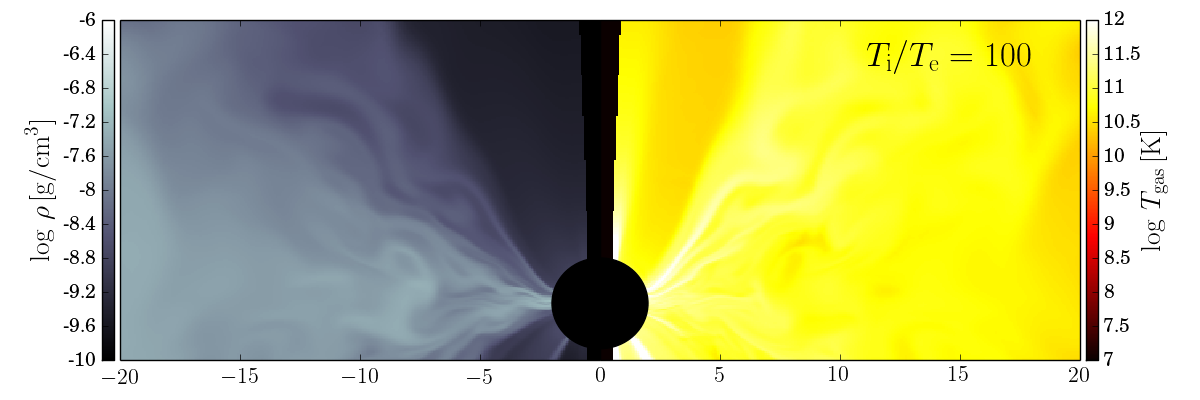}\vspace{-.5cm}
  \includegraphics[width=1.5\columnwidth]{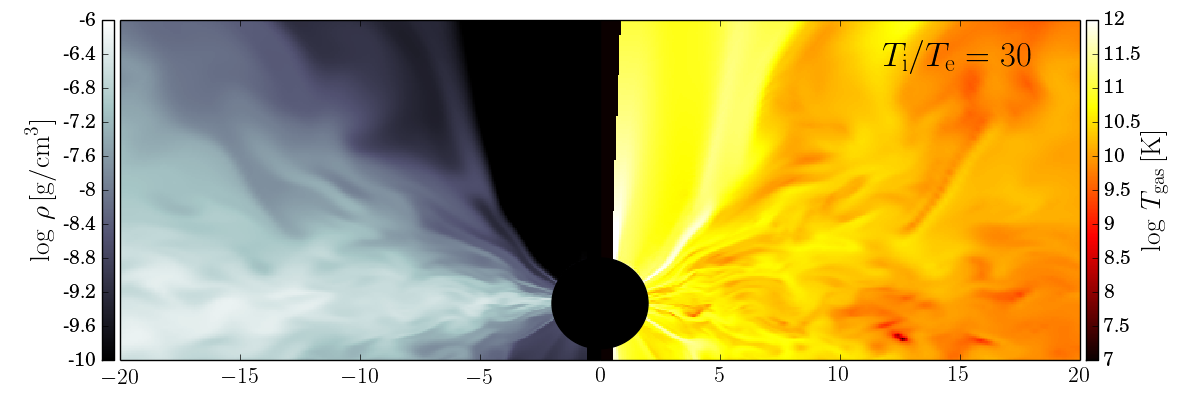}\vspace{-.5cm}
 \includegraphics[width=1.5\columnwidth]{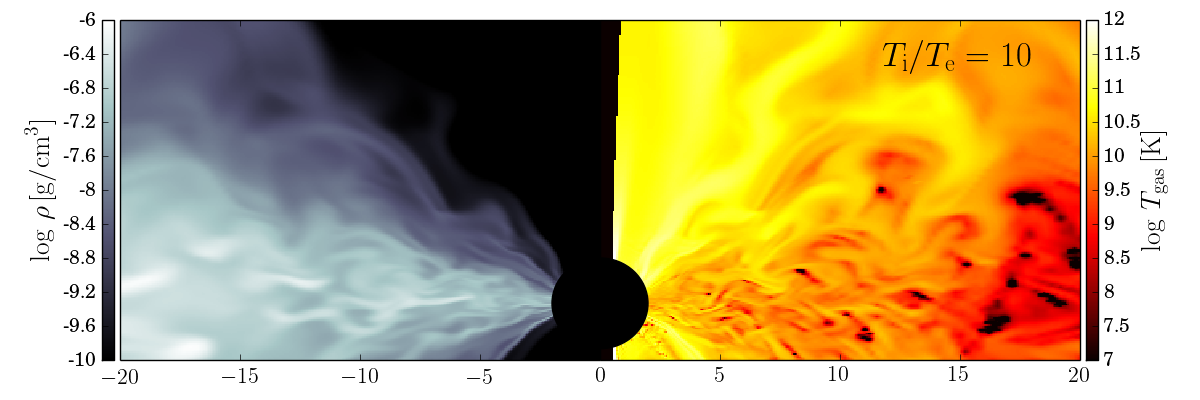}
\caption{Distributions of gas density (left panels) and gas temperature (right panels) in simulated accretion flows
corresponding roughly to the accretion rate of $10^{-3}\Medd$ and three values of ion to electron temperature ratio: (top to bottom)  $T_{\rm
  i}/T_{\rm e}=100$ (least-efficient, model \texttt{f5t100}), $30$ (\texttt{f5t30}), and $10$ (most-efficient  radiative cooling, \texttt{f5t10}).}
  \label{f.rhotemp}
 \end{figure*}

 The simulations initialized as described in the previous section
   evolve into turbulent, magnetorotational instability (MRI) driven,
   optically thin accretion flows. The inflow of matter takes place in
   the bulk of the disk where turbulent stresses take angular momentum
   out of the gas. The bulk of the flow is relatively strongly
   magnetized due to the non-zero radial component of the initial
   magnetic field at the equatorial plane, with magnetic pressure
   contributing ultimately to ${\sim 30\%}$ of the total pressure in
   the region $10<R/\Rg<20$. The magnetocentrifugal forces drive
   outflow from the surface layers of the disk. Due to its low density, 
   the outflowing gas does not
   contribute significantly to the total radiative emission.

 Different densities at the onset of the simulations determined the
 efficiency of radiative cooling -- the larger the density, the higher
 the relative bremsstrahlung cooling rate. Similarly, the larger the
 electron temperature for a given gas temperature, the more efficient
 is the cooling. The latter effect is clearly visible in
 Fig.~\ref{f.rhotemp} which shows the poloidal slice of the gas
 density (left panels) and temperature (right panels) for three models
 with roughly the same accretion rate of a few $10^{-3}\Medd$:
 \texttt{f5t100}, \texttt{f5t30}, and \texttt{f5t10}. While the
 densities are of the same order, the temperatures are not.  The
 simulation with the coldest electrons (\texttt{f5t100}, $T_{\rm
   i}/T_{\rm e}=100$, top panel) is as hot as it gets with gas
 temperature close to the virial temperature, as expected for a
 non-radiative accretion flow. As soon as electrons get hotter
 relative to ions (model \texttt{f5t30}, middle panel), the radiative
 losses are significant enough to affect the gas properties -- the gas
 temperature (the mean of electron and ion temperatures) noticeably
 decreases. Gas at the equatorial plane, at $R=15\rg$ is now at $\sim
 10^{10}\rm\,K$, compared to $\sim 10^{11}\rm\,K$ for model
 \texttt{f5t100}. Increasing the electron relative temperature
 increases further the cooling effect. For the lowest ion-to-electron
 temperature ratio ($T_{\rm i}/T_{\rm e}=10$, bottom panel, model
 \texttt{f5t10}), the gas temperature at this location falls down to
 $\sim 10^{9}\rm\,K$. In addition to the gas temperature, the
 radiative cooling have an effect on the disk thickness - the two runs
 affected by cooling have noticeably smaller thickness (see Table~\ref{t.models}), but they are far 
from collapsing to a thin disk.\footnote{Their properties resemble to 
large extent the luminous hot accretion flow (LHAF) solutions proposed
by \cite{yuan-lhaf}. Detailed study of the collapse would require precise 
treatment of Coulomb coupling (see Discussion).
} At the lowest accretion rates, the radiative emission is dominated by
synchrotron, while at the largest, for which the radiative cooling has significant
impact on gas properties, it is predominantly Compton cooling (contributing, e.g., to ${\sim}95\%$ of cooling at the equatorial plane for model \texttt{f5t30}) and, to lesser extent, bremsstrahlung \citep[see, e.g.][]{narayan-yi95, esin+97}.

\subsection{Mass and energy transfer rates}

Quasi-stationary, i.e., accreting at on average constant rate,
accretion flows show mass and total energy transport rates independent
of radius. Under such condition, neither gas or energy accumulates at
any radius, but preserves their average radial profiles. We measure
the accretion rate and luminosity in total energy in simulated
accretion flows by integrating the mass and
energy fluxes over the BH horizon.

The mean accretion rate is therefore defined through,
\be \label{e.mdot} 
\dot M = \int_{0}^\pi
\int_0^{2\pi}\sqrt{-g}\,\langle\rho u^r\rangle{\rm d}\phi {\rm d}\theta,
\ee
where $\rho$ stands for gas density, $u^r$ denotes the radial velocity, and $\sqrt{-g}$ is the metric determinant. 

The luminosity in total energy (binding, kinetic, radiative, magnetic, and thermal, see \citealp{sadowski+enfluxes} for the related discussion) is given as,
\be \label{e.entot} 
L_{\rm tot} = -\int_{0}^\pi
\int_0^{2\pi}\sqrt{-g}\,\langle T^r_t + R^r_t+\rho u^r\rangle{\rm d}\phi {\rm d}\theta,
\ee
where $T^r_t$ and $R^r_t$ are components of the matter and radiation stress energy tensors, respectively.

The radiative-only luminosity is given by the integral of the latter,
\be \label{e.enrad} 
L_{\rm rad} = -\int_{0}^\pi
\int_0^{2\pi}\sqrt{-g}\,\langle R^r_t\rangle{\rm d}\phi {\rm d}\theta.
\ee
However, because radiative energy is not conserved (radiation can be generated and absorbed by the gas), we need to carefully choose the radius at which to perform the integration. Ideally, we would like to measure the luminosity at infinity. However, we are limited by the region where the accretion flow has settled to the converged solution. We decided to perform the integrals at $R=20\,\rg$, which is not infinity, but encompassed the region where most of the radiation in optically thin disks is formed.

Assuming that the energy liberated in an accreting system can ultimately escape only either by radiation or mechanical energy, we can calculate the kinetic component as,
\be
\label{e.Lkin}
L_{\rm kin}=L_{\rm tot}-L_{\rm rad}.
\ee

\subsection{Radiative and kinetic luminosities}

The efficiency associated with given luminosity is defined as,
\be
\eta=\frac{L}{\dot M c^2},
\ee
where $\dot M$ is the accretion rate through the BH horizon and
$c$ is the speed of light.

\begin{figure}
  \includegraphics[width=1.\columnwidth]{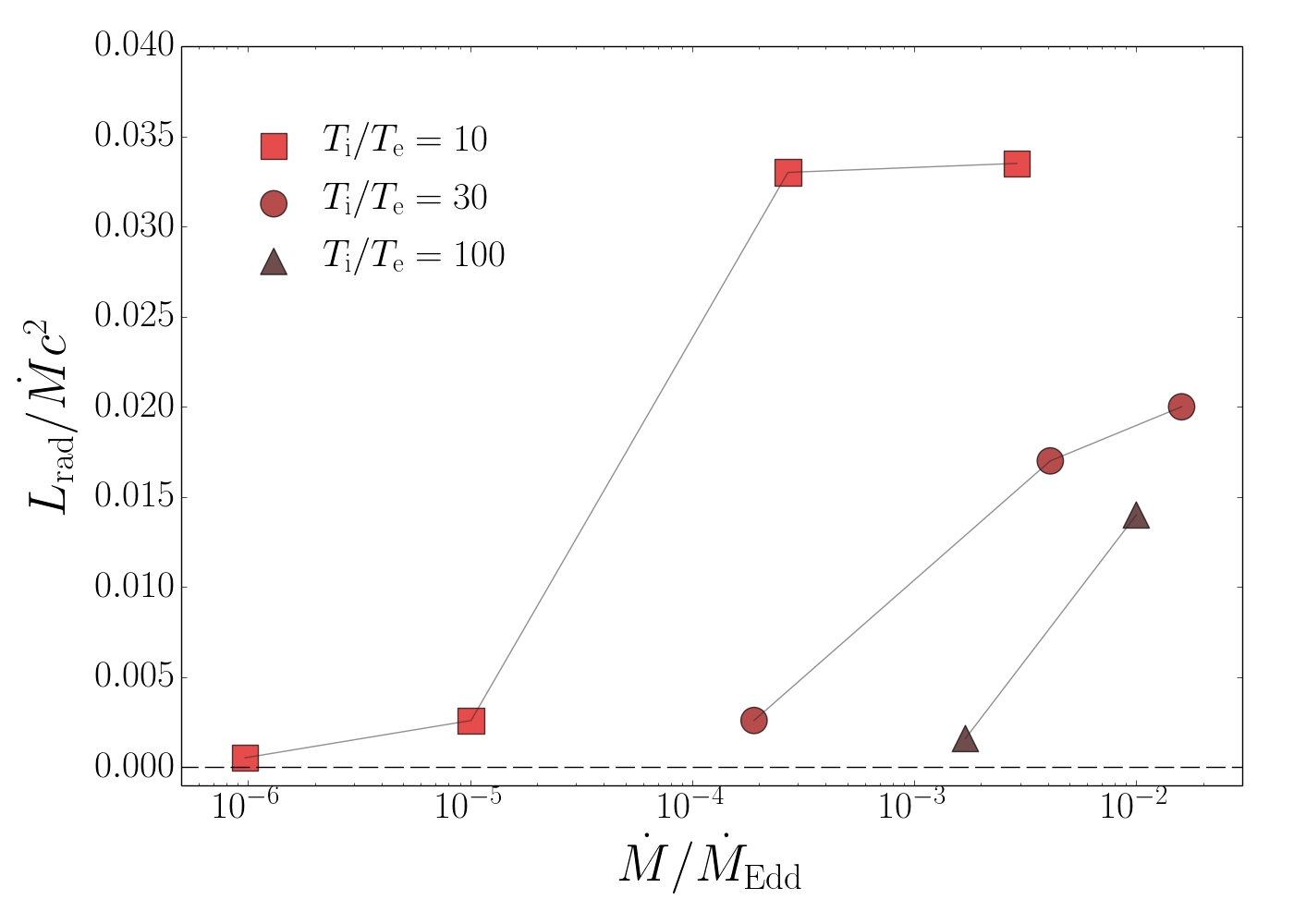}\vspace{-.65cm}
 \includegraphics[width=1.\columnwidth]{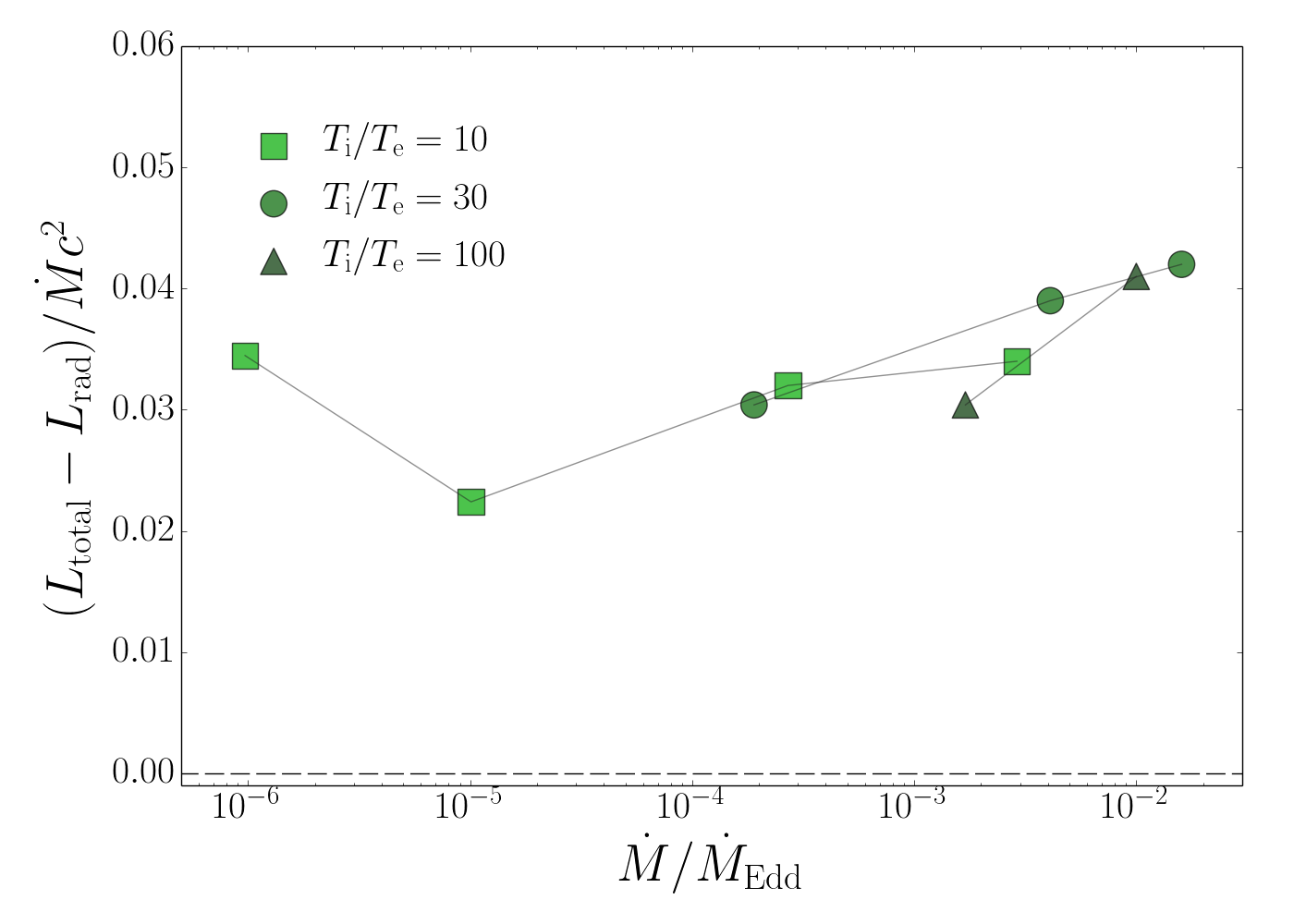}
\caption{Radiative (top) and kinetic (bottom panel) efficiencies of the simulated accretion flows as a function of the normalized accretion rate. Different symbols correspond do different ion to electron temperature ratios.}
  \label{f.effs}
 \end{figure}

\begin{figure}
  \includegraphics[width=1.\columnwidth]{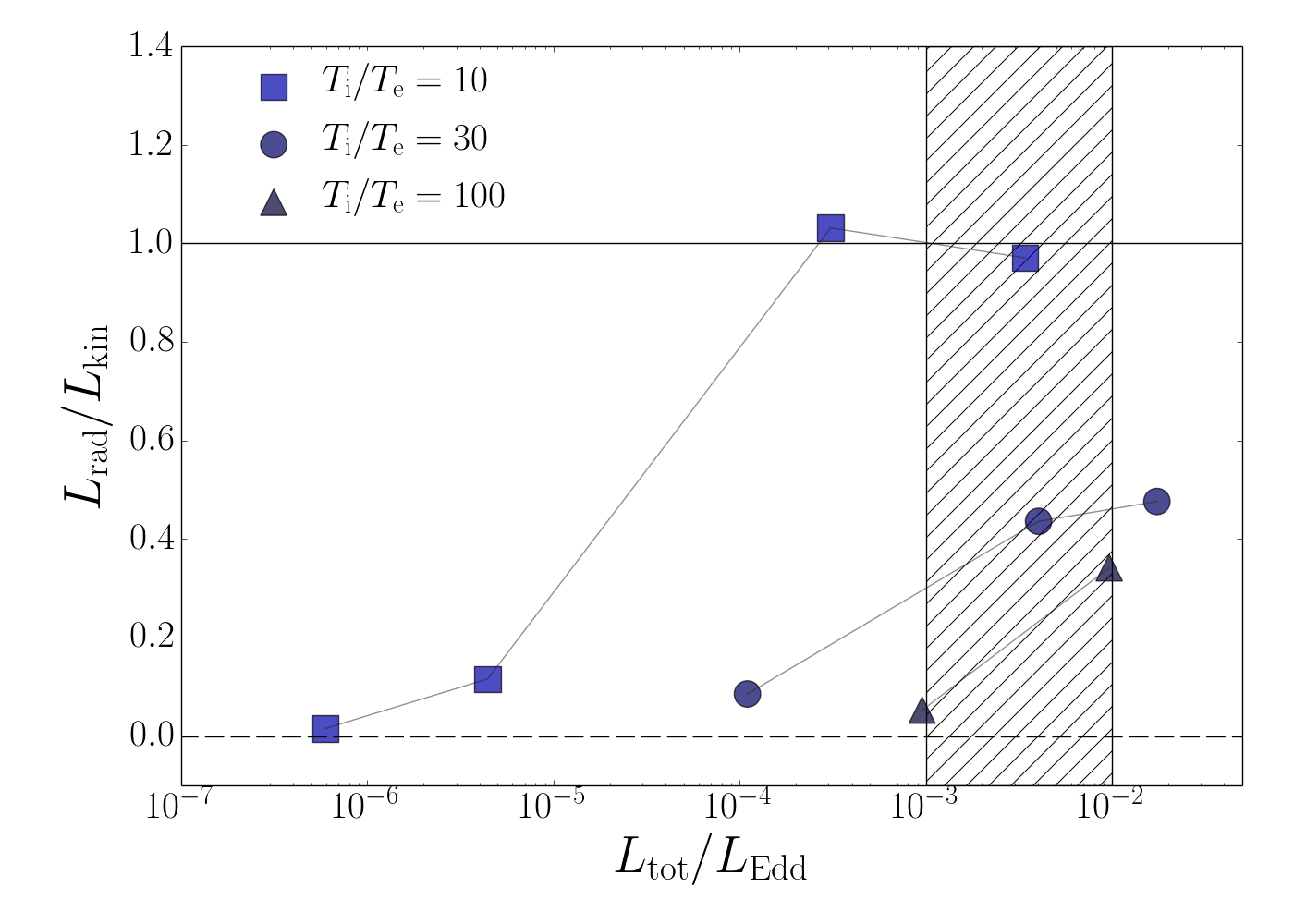}
  \caption{Ratio of the radiative to kinetic luminosities as a
    function of the total luminosity (sum of the two). The horizontal line corresponds to the two
    equal in value and separates the radiatively inefficient regime
    (below) from the luminous state (above the line). The 
shaded region reflects the range of luminosities 
  at which the observed AGN start
    producing radiative and kinetic outputs of comparable magnitude (Russell et
    al.~2013).}
  \label{f.effsratio}
 \end{figure}

Table~\ref{t.models} lists the nine simulations we performed and gives the 
accretion rates and efficiencies obtained by applying formulae from the previous
section to time- and azimuth-averaged data spanning the last $8000\,\tg$ of
each simulation. These efficiencies are presented graphically in Figs.~\ref{f.effs} and \ref{f.effsratio}.

The top panel in Fig.~\ref{f.effs} shows the radiative efficiencies as
a function of the accretion rate normalized to the Eddington value
(Section~\ref{s.units}). The squares, circles and triangles correspond
to ion to temperature ratios of $T_{\rm i}/T_{\rm e}=10$ (hottest
electrons), $30$, and $100$ (coldest electrons), respectively.  Within
each subgroup, the radiative efficiency increases with accretion rate
-- growing gas density results in increased efficiency of
bremsstrahlung emission. The accretion rate at which an accretion flow
departs from the radiatively inefficient state ($\eta_{\rm rad}\ll 1$)
is different for different electron temperatures. For the hottest
electrons case ($T_{\rm i}/T_{\rm e}=10$, square points), already at
$\sim 10^{-4}\Medd$ (model \texttt{f4t10}) the amount of the produced
radiation accounts to $3\%$ of the accreted rest-mass energy - 
roughly the amount that the standard, radiatively efficient
thin disk \citep{ss73,nt73} would generate inside radius $20\rg$. For colder ellectrons, this transition
takes place at significantly higher accretion rates: $\sim
10^{-3}\Medd$ for $T_{\rm i}/T_{\rm e}=30$, $\sim
10^{-3}-10^{-2}\Medd$ for $T_{\rm i}/T_{\rm e}=100$. In those cases,
the radiative efficiency does not exceed $2\%$ even for the highest
simulated accretion rates. Such a discrepancy in the critical
accretion rate corresponding to the transition to the luminous regime
is not unexpected -- the hotter the electrons (with respect to gas or
ions), the more efficient is radiative emission, and lower gas
densities (and corresponding accretion rates) are required for the
same amount of emission. Comparable results were obtained
by \cite{xieyuan-12} who calculated radiative efficiencies using semi-analytical, one-dimensional solutions parametrized, however, by the electron heating parameter, $\delta$, not the temperature ratio. Similar calculations were also performed in the context of M87 by \cite{ryan-bhlight} who simulated GRMHD accretion flows affected by radiative cooling effects tracked with a Monte Carlo radiative transport solver. For equal ion and electron temperatures ($T_{\rm i}/T_{\rm e}=1$) and accretion rate ${\sim}10^{-5}\dot M_{\rm Edd}$ they obtained radiative efficiency $L_{\rm rad}=0.51\dot M c^2$ -- as authors note, surprisingly high even for their choice of BH spin value, $a_*=0.9375$, and presumably due to the flow having not yet reached steady state\footnote{Our own experiments have shown that axisymmetric accretion flows are characterized by significantly higher radiative efficiency than the three-dimensional counterparts  -- enforcing axisymmetry changes the nature of turbulence leading to the formation of rapidly radiating, high density filaments in the inner region which are not present in non-axisymmetrical solutions.}. Numerical simulations of the transitional regime were performed, although with much simpler treatment of radiation, also by \cite{dibi+12} and \cite{wu+16}.

The bottom panel in Fig.~\ref{f.effs} shows the kinetic efficiency calculated from the difference between the total and
radiative luminosities (Eq.~\ref{e.Lkin}). In all cases the kinetic efficiency is close to 3\%.  This value is consistent with that observed in the non-radiative simulation described in-depth  in \cite{sadowski+enfluxes}\footnote{We remind this 3\% characterizes thick accretion flows on non-rotating BHs. For non-zero BH spin this disk-related efficiency may go up; on top of it, a generated relativistic jet may overcome the total energy budget, albeit having difficulty to efficiently couple with the host halo due to the very high collimation.}. 
It is interesting to note, that the kinetic efficiency does not go down with increasing accretion rate and radiative luminosity.
This is in part different from previous idealized AGN models \citep{churazov+05}. It reflects the fact that radiation comes from the innermost part
of the accretion flow and the underlying gas cannot efficiently respond to cooling in the last phase of accretion, crossing the horizon with intact dynamical properties (the total efficiency is determined by how bound such horizon crossing gas is).
For the lowest accretion rates in each subset, this kinetic energy extraction rate dominates the total luminosity. In the case of the largest accretion rates, for which the radiative output is significant, the kinetic and radiative luminosities have comparable magnitudes.

The kinetic and radiative energy extraction rates are compared in  Fig.~\ref{f.effsratio}. As already discussed, the kinetic energy dominates when radiative emission is negligible, i.e., for the lowest accretion rates. Once mass transfer rate increases, so does the radiative efficiency, which for the hottest electron case can even become comparable or exceed the kinetic component.

Table~\ref{t.transition} puts together the estimates for the luminosities, $L_{\rm trans}$ (either kinetic or radiative), at which the transition to the luminous regime ($L_{\rm rad}\approx L_{\rm kin}$) takes place. For the coldest electrons, we have shown that the accretion flows remain radiatively inefficient up to $10^{-2}L_{\rm Edd}$ and the radiative output never reaches the level of the kinetic one. The intermediate case behaves in similar way. For the hottest electron case ($T_{\rm i}/T_{\rm e}=10$), however, the accretion flow starts generating radiative luminosity comparable to the mechanical one already at $10^{-4}L_{\rm Edd}$. We discuss the implications of these findings in the next Section.

\begin{table}
\begin{center}
\caption{Transition to the luminous state}
\label{t.transition}
\begin{tabular}{lc}
\hline
\hline
$T_{\rm i}/T_{\rm e}$ & $L_{\rm trans}/L_{\rm Edd}$ \\
\hline
\hline
$100$ & $\gtrsim 10^{-2}$ \\
$30$ & $\gtrsim 10^{-2}$ \\
$10$ & $\sim 10^{-4}$ \\
\hline
\hline
\multicolumn{2}{l}{Luminous state defined as }\\
\multicolumn{2}{l}{satisfying $L_{\rm rad}>L_{\rm kin}$.}\\
\end{tabular}
\end{center}
\end{table}

\section{Implications}
\label{s.implications}

Observational studies of AGN radiative luminosities and mechanical output (Section \ref{s.constraints}) have shown that the radiative component becomes comparable with the kinetic one at the luminosity $\sim 10^{-3}-10^{-2} L_{\rm Edd}$ \citep[e.g., Fig.~12 in][]{russell+13}. In this work we have shown that this critical luminosity is sensitive to the temperature of the electrons, which determines the efficiency of radiative cooling. Therefore, we can use the observational result to constrain the properties of the electron population in optically thin black hole accretion flows.

In Table~\ref{t.transition} we listed the rough estimates of the critical luminosities at which the kinetic and radiative luminosities become comparable. For the hottest electrons case we considered (ion to electron temperature ratio, $T_{\rm i}/T_{\rm e}=10$), this transition occurred for luminosity as low as $10^{-4}L_{\rm Edd}$ (see Fig.~\ref{f.effsratio}), significantly below the observational constraint. For both the intermediate ($T_{\rm i}/T_{\rm e}=30$) and the coldest ($T_{\rm i}/T_{\rm e}=100$) cases, the radiative output is significantly below the mechanical one even for kinetic luminosities $\sim10^{-2}L_{\rm Edd}$, well above the constraint.

This argument implies that to satisfy the observational constraints one should expect the temperature ratio to fall between the hottest and intermediate case, i.e., $10<T_{\rm i}/T_{\rm e}<30$. We therefore suggest that astrophysical plasmas in optically thin accretion flows (at least in the nuclear region) produce electrons with temperature between 3\% and 10\% of the ion temperature.

The physics of electron heating in accretion flows depends on the
phenomena taking place on the smallest lengthscales of collisionsless
plasmas \citep[see, e.g.,][]{quataertgruzinov-99} and is still
debated. \cite{howes-10} proposed a prescription for the fraction of
heating going into the electron population based on theoretical models
of the dissipation of MHD turbulence in almost collisionless plasmas
and which have been recently applied to simulations of two-temperature
accretion flows \citep{ressler+15,sadowski+electrons}. This model
predicts that the efficiency of electron heating increases in highly
magnetized regions, and the equilibrium species temperature ratio is
determined by the magnetization of local plasma. Putting the range of
temperature ratios favored by this work, $10<T_{\rm i}/T_{\rm e}<30$,
into the formula for the equilibrium temperature ratio
\citep{sadowski+electrons}, one recovers the corresponding range of
gas-to-magnetic pressure ratio, $6\lesssim \beta\lesssim 10$. This
level of magnetization is in agreement with the level expected from
the saturated state of magnetorotational instability
\citep[e.g.,][]{davis+10,shi+10}, what further supports the arguments
made, and suggests that the accretion flows of interest are not
threaded by significant net poloidal magnetic fluxes which would imply
much stronger magnetization \citep{xuening+13,salvesen+16}.

A key result worth to remark is the robustness of the kinetic efficiency, which remains stable around 3 percent value regardless of the Eddington ratio, at most varying by 1 percent. While broadly accepted that the maintenance mode of AGN feedback resides in the kinetic regime, given the ubiquitous imprints of AGN outflows via X-ray cavities, shocks, and gas uplift (Section \ref{s.intro}), the kinetic output in the quasar regime is still debated\footnote{We note that even in the quasar-like regime the radiative power has severe difficulty in coupling with the gas, especially at $>$\,kpc scales.}. Previous models assumed a strong decrease of the kinetic power at increasing Eddington ratios (e.g., \citealp{churazov+05}), however recent observational data has started to detect the presence of cavities even in systems with a quasar source. E.g., \cite{mcdonald+15} show the presence of X-ray cavities in Phoenix cluster which also hosts an X-ray bright quasar emitting at a few percent of the Eddington rate.
Albeit the very high Eddington ratios remain to be tested, our work corroborates the importance of mechanical AGN feedback also in the transition to the quasar-like regime, and thus even at high redshifts (as suggested by the observational sample in \citealp{hlavacek+15}). 

Finally, as anticipated in Section \ref{s.intro}, the robust constraint and stability of the mechanical efficiency allows us to link such micro value to the macro properties of the self-regulated AGN feedback loop (mediated via CCA), which re-heats and preserves the cooling cores of galaxies, groups, and clusters in a state of quasi-thermal equilibrium throughout the cosmic time (e.g., \citealp{gaspari15} and refs. within). 
Such unification model is presented and discussed in-depth in the companion work, \cite{gasparisadowski+17}.

\section{Caveats}
\label{s.caveats}

We base our work on simulations performed with a numerical method
using some simplifying assumptions. Most importantly, we decided not
to track the energy contained in the electron fluid independently of
ions, but to fix the ion to electron temperature ratio. This approach
effectively arbitrarily prescribes the balance between the electron
heating and Coloumb coupling which determines the temperature
ratio. We adopted this simplification to get a direct handle on the
electron temperatures and to avoid numerical issues that arise when
identifying dissipation for the purpose of electron evolution
\citep[see discussion in][]{sadowski+electrons}. The trade-off is that
we force the temperature ratio to be spatially uniform -- which is not
the case in the whole volume, but is close to constant inside the bulk
of the accretion flow \citep{ressler+15,
  sadowski+electrons}. Furthermore, we do not track the efficiency of
Coulomb coupling, which at the largest accretion rates (and densities)
can couple electrons and ions strongly enough to prevent 
large discrepancy 
between their temperatures. On the other hand, if a given temperature
ratio is feasible, then the above results apply, as the radiative
properties depend ultimately on the electron temperature, and not on
the processes which determine its value.

We limited our set of simulations to accretion flows on non-rotating
BHs. If the BH spin is non-zero, one may expect higher efficiency of
extracting both radiative and mechanical energy
\citep{sasha+madjets,mckinney+radmad,sadowski+enfluxes}, and it is
likely the transition luminosities identified in this work will
change\footnote{Another complication is that it is uncertain how
efficiently the chaotic inflow would align with the BH spin axis.}. 
However, if BH growth occurs mainly through chaotic accretion,
as expected for most massive galaxies, groups, and clusters
\citep{gaspari+13,gaspari+15,gaspari+17}, one should expect the
average value of the BH spin in AGN to be close to zero
\citep{king+06}. In particular, a sample as in \cite{russell+13}
should be biased toward the properties of a zero-spin BH population.

It has been understood recently that the level of magnetization and
the topology of magnetic field in an accretion flow are not unique,
and depend on the large scale properties of the field. Magnetic field
may be considered another degree of freedom in determining the flow
properties, which we leave to future work. By choosing a particular
configuration of the initial seed magnetic field in the equilibrium
torus, we chose to study accretion flows which are relatively weakly
magnetized, and which do not lead to the saturation of the field at
the BH horizon. The opposite limit would be the magnetically arrested
disk \citep[MAD,][]{narayan+mad}, which shows properties dramatically
different from the weakly magnetized state \citep[SANE -- using the
nomenclature in][]{narayan+adafs}. Whether low luminosity AGN
accretion flows are SANE or MAD is still debated. In this work we
focused only on the SANE scenario.

\section{Summary}
\label{s.summary}

We performed a set of general relativistic, radiative, magneto-hydrodynamical simulations to study the transition from radiatively inefficient to efficient state of accretion on a non-rotating BH. We studied a range of ion to electron temperature ratios, and simulated flows corresponding to accretion rates as low as $10^{-6}\Medd$, and as high as $10^{-2}\Medd$. We have found the following:

(i) The radiative efficiency increases with accretion rate, and the
increase occurs earlier for lower ion to electron temperature ratios
(i.e., hotter electrons). For $T_{\rm i}/T_{\rm e}=10$, the radiative
output becomes significant (exceeding 1\% of the rest mass energy
flux) already at $10^{-5}$-$10^{-4}\Medd$. For colder electrons, this
transition takes place only at ${\sim}10^{-3}$ and
${\sim}10^{-2}\Medd$ for $T_{\rm i}/T_{\rm e}=30$ and $T_{\rm
  i}/T_{\rm e}=100$, respectively (Fig.~\ref{f.effs}).

(ii) The mechanical output of the accretion flows is insensitive to the accretion rate, with an efficiency remaining stable near $3\%$ of $\dot M c^2$. Such micro efficiency can be exploited to close a unified self-regulated AGN feedback model which shapes massive galaxies, groups, and clusters (see the companion \citealp{gasparisadowski+17}). We remark this is also valid as entering the quasar regime, corroborating the importance of mechanical AGN feedback over a large range of Eddington ratios (and thus redshifts).

(iii) By comparing the simulated radiative and mechanical efficiencies with the observational results constraining the luminosity at which the radiative output becomes comparable with the mechanical power \citep{russell+13}, we show that our models are consistent with observations when the ion to electron temperature ratio is within a range $10<T_{\rm i}/T_{\rm e}<30$.

\section{Acknowledgments} 

The authors thank Sean Ressler for comments on the manuscript. AS and MG acknowledge support
for this work 
by NASA through Einstein Postdoctotral Fellowships number PF4-150126 and PF5-160137, respectively,
awarded by the Chandra X-ray Center, which is operated by the
Smithsonian
Astrophysical Observatory for NASA under contract NAS8-03060. 
Support for this work was also provided by NASA Chandra award number G07-18121X.
The authors acknowledge computational support from NSF via XSEDE resources
(grant TG-AST080026N), from the PL-Grid Infrastructure, and the NASA/Ames HEC Program (SMD-16-7251). 

\nocite{*}

\bibliographystyle{apj}

%\bibliography{biblio.bib}

\end{document}